\documentclass[prl,twocolumn,floats,showpacs,superscriptaddress]{revtex4}
\usepackage{psfig}
\begin{document}

\title{Stochastic theory of synchronization transitions in extended systems}

\author{Miguel A. Mu{\~n}oz} 

\affiliation{\mbox{Instituto de F{\'\i}sica Te{\'o}rica y Computacional Carlos I,
  Facultad de Ciencias, Univ. de Granada, 18071 Granada, Spain}}

\author{Romualdo Pastor-Satorras}

\affiliation{\mbox{Departament de F{\'\i}sica i Enginyeria Nuclear Universitat
  Polit{\`e}cnica de Catalunya Campus Nord, 08034 Barcelona,
  Spain}}

\date{\today}

\begin{abstract} 
  We propose a general Langevin equation describing the universal
  properties of synchronization transitions in extended systems.  By
  means of theoretical arguments and numerical simulations we show
  that the proposed equation exhibits, depending on parameter values,
  either: i) a continuous transition in the bounded
  Kardar-Parisi-Zhang universality class, with a zero largest Lyapunov
  exponent at the critical point; ii) a continuous transition in the
  directed percolation class, with a negative Lyapunov exponent, or
  iii) a discontinuous transition (that is argued to be possibly just a 
  transient effect).  Cases ii) and iii)
  exhibit coexistence of synchronized and unsynchronized phases
  in a broad (fuzzy) region. This phenomenology reproduces almost all the 
  reported features of synchronization transitions of coupled map 
  lattices and other models, providing a unified theoretical framework 
  for the analysis of synchronization transitions in extended systems.
\end{abstract} 
\maketitle 

Arrays of coupled oscillators appear ubiquitously in Nature, and their
mutual synchronization is undoubtedly one of the most intriguing
phenomena appearing in complex systems \cite{Strogatz-Syn}. Some
well-known instances of synchronizing systems are chemical reactions
\cite{chemical}, neuronal networks \cite{neurons}, flashing fireflies
\cite{fireflies}, Josephson junctions \cite{Josephson}, electronic
circuits \cite{circuits}, and semiconductor lasers \cite{lasers}. In
particular, coupled map lattices (CMLs) \cite{Kaneko}, initially
introduced as simple models of spatio-temporal chaos \cite{CM}, have
received a great deal of attention, as models of synchronization in
spatially extended systems (another possibility is to analyze discrete
cellular automata \cite{Peter,death}).  In this context, it was
observed that distant patches of a given CML can oscillate in phase
\cite{CM}, a phenomenon related to the Kardar-Parisi-Zhang (KPZ)
equation \cite{Geoff} describing the roughening of growing interfaces
\cite{HZ}.  Afterwards, it was also realized that when two different
replicas of a same CML are {\it locally} coupled, they become
synchronized if the coupling strength is large enough
\cite{Peter,Zanette,circuits}.  Also {\it globally} coupled CML can
achieve mutual synchronization \cite{Kaneko2} with
interesting implications in neuro-science \cite{ZM}.  Last but not
least, different replicas of a CML can be synchronized if they are
coupled to a sufficiently large common external random noise, even if
they are not directly coupled to each other
\cite{firenze1,first,lucia}.

In all these examples, there is a transition from a chaotic or
unsynchronized phase in which perturbations grow, and two replicas
evolve independently, to a synchronized phase in which memory of the
initial difference is asymptotically lost and replicas synchronize.
This synchronization transition (ST) resembles very much other 
non-equilibrium critical phenomena, particularly transitions into 
{\it absorbing states} \cite{HGM}, and the determination of its universal
properties has become the subject of many recent studies
\cite{Peter,lucia,PK,Ahlers} as well as the main motivation of this
Letter.  Within this context, a major contribution was made by
Pikovsky and Kurths (PK) who proposed a stochastic model for the
dynamics of perturbations in locally coupled CML, in which, under very
general conditions, the difference-between-two-replicas field can be
described by the so-called multiplicative noise (MN) Langevin equation
\cite{PK}:
\begin{equation}
   {\partial}_{t} \phi(x,t) = - a \phi -
   b \phi^2 + D {\nabla}^{2} \phi(x,t) +  \sigma  \phi(x,t) \eta,
\label{MN}
\end{equation}
where $\phi(x,t)$ is the difference field (or ``synchronization error''),
$a$, $b$, $\sigma^2$, and $D > 0$  are parameters,
and $\eta$ a delta correlated Gaussian white noise with zero average.
It is worth remarking that the non-linear saturation term was added in
\cite{PK} by hand, in order to ensure that the synchronization error
remains bounded from above (i.e. of the order of the original field
variables) \cite{PK,Ahlers}.

Eq. (\ref{MN}) has been extensively studied, unveiling a rich
phenomenology, including a novel non-equilibrium phase transition
\cite{MN}.  This equation is related to: i) the problem of directed 
polymers in random media
\cite{PK,HZ} and, ii) the KPZ equation in the presence of a limiting
upper wall \cite{MN}.  To verify this last, particularly illuminating
relation, it is useful to perform a Cole-Hopf transformation
$\phi(x,t)=\exp[h(x,t)]$ \cite{MN,Ahlers}, obtaining
\begin{equation}
\partial_t h(x,t) = -a - 
b ~\exp{h} + D \nabla^2 h + D (\nabla h)^2 + \sigma \eta,
\label{KPZ}
\end{equation}
which is indeed a KPZ equation plus a bounding term---the upper
wall---pushing positive values of the height $h$ towards negative
values \cite{PK,MN}. The synchronized or absorbing phase corresponds
in this language to a depinned interface, falling continuously towards
minus infinity, governed by KPZ dynamics. The unsynchronized or
active phase, on the other hand, translates into a pinned-to-the-wall
interface.

Using the PK theory, it could be naively concluded that all ST belong
to the MN universality class \cite{PK}.  On the other hand, it has
been known for some time that many ST in discrete cellular automata
(CA) belong generically to the directed percolation (DP)
universality class \cite{Peter}.  This difference led Grassberger to
conjecture that the discreteness of the local field variables is the
factor that determines a ST universality class: MN for continuous
variables and DP for discrete ones \cite{Peter}. Following
Grassberger, the key difference lies in the fact that while for
continuous variables one has `incomplete death' (fluctuations can
always generate unsynchronized regions from synchronized ones
\cite{death}), discrete systems do not allow such a process (`complete
death'). This difference between MN and DP has been characterized in
Ref.~\cite{nature}.

Some counterexamples, however, have recently shown that this
conjecture does not apply all the way: in some cases, CML (continuous
variables) can exhibit DP behavior \cite{lucia,Ahlers}, or present a
discontinuity at the transition point \cite{first}.  Based on these
observations a new global picture has emerged: there are two types of
ST depending on whether the largest Lyapunov exponent, $\Lambda$, is zero
or negative at the transition.  In the first case, the transition is
MN-like and controlled by linear stability effects ($\Lambda$ changes sign
at the transition).  In the second case the transition is either DP or
discontinuous, depending on microscopic details, and the transition is
located where the propagation velocity of the non-linear perturbations
changes sign, while $\Lambda$ is negative.

The aim of this Letter is to construct a stochastic model, including Eq.
(\ref{MN}) as a particular case, able to reproduce in a unified
framework all the aforementioned phenomenology. With this purpose, we
consider the following Langevin equation
\begin{eqnarray}
{\partial}_{t} \phi(x,t) & = & - \frac{\delta V(\phi)}{\delta \phi} 
+ D {\nabla}^{2} \phi(x,t) +  \phi(x,t) \eta(x,t),  \nonumber \\
V(\phi) &= & a \phi^2 + b \phi^3 + c \phi^4,
\label{Wet}
\end{eqnarray}
which has appeared in the literature in the context of
non-equilibrium wetting and depinning transitions
\cite{haye,Lisboa-wet}. This is a generalization of Eq. (\ref{MN})
including saturation effects in a more general way.  In particular,
for $b>0$, the parameter $c$ is irrelevant in the renormalization
group sense, and Eq.~(\ref{Wet}) behaves as Eq.~(\ref{MN}) (i.e.  it
belongs to the MN universality class).  For $b<0$, however, new
physics emerges \cite{Lisboa-wet} (note that in the KPZ language $b<0$ 
translates into the presence of an {\it attractive} upper wall). 
This second possibility is required to describe transitions in which a 
non-vanishing value of the synchronization-error field is a local 
attractor of the dynamics.  
For example, for discrete boolean CA, the value $\phi(x)=1$ should be 
a minimum of the local deterministic dynamics, and therefore $b<0$ is 
required to describe them within this formalism. Also, for discontinuous CML
\cite{lucia,Ahlers} this term describes finite jumps from one
map-sector to another.

\begin{figure}
\centerline{\psfig{file=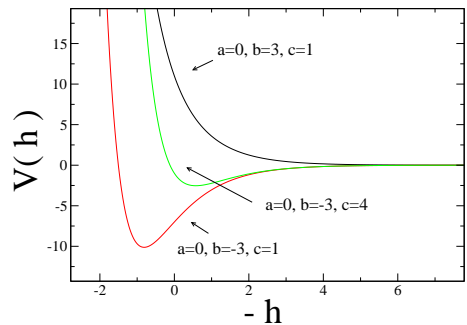,width=6cm,angle=-90}}
\caption{Effective potential in the 0-dimensional case in terms of  $-h$
 for values of $a$ in the depinned phase.  Note the attractive effect
 for $b<0$.}
\label{Fig1}
\end{figure}

In Fig~\ref{Fig1} we show a plot of the exact single-site effective
potential in the zero-dimensional limit in the interface representation 
(i.e. the stationary solution of the associated Fokker-Planck 
equation after performing the Cole-Hopf transformation; see Ref.~\cite{nature}).  
For $b<0$ a new minimum (absent for $b>0$), 
present even in the synchronized phase, 
appears---the KPZ wall becomes attractive.  
As long as $b > 0$, $c$ is irrelevant \cite{ir},
the transition is in the MN universality class, and it occurs where
the depinned interface looses its stability, implying that the average
velocity $ \langle \dot{h} \rangle$ changes sign at the transition point. It is
also well known \cite{PK} that $\langle \dot{h} \rangle$ coincides exactly with
the largest Lyapunov exponent, $\Lambda$. Therefore, for $b>0$, the
situation is completely analogous to ST in the MN class where $\Lambda$
changes sign at the transition. On the other hand, for $b<0$
\cite{ir} the transition can be easily shown not to occur at $\langle
\dot{h} \rangle=0$ \cite{Lisboa-wet}. Instead, there is a finite interval
of values of $a$, $[a^*,a_c(b)]$, in which phase coexistence is found,
ergodicity is broken, and the final state depends upon initial
conditions \cite{Lisboa-wet}. Within this interval, pinned interfaces
remain usually pinned, and depinned interfaces escape towards minus
infinity at a finite velocity.  Below (above) $a=a^*$ ($a_c(b)$) depinned
(pinned) interfaces lose their stability. Consequently, at $a_c$ the
depinned phase is stable (contrary to what happens for $b>0$) and $\langle
\dot{h}(a_c) \rangle=\Lambda$ is negative, in analogy with ST in either the DP
class or discontinuous transitions   where $\Lambda<0$ 
(as explained afterwards the
first-order transition could be a transient effect).
 
In order for Eq.~(\ref{Wet}) to constitute a valid theory for ST, a DP
critical point should be observed in some parameter range for $b<0$.
To the best of our knowledge DP behavior have not been reported for
this equation so far. As we will show,
DP behavior may emerge for $b<0$ as the forthcoming
numerical results show.

To simulate Eq.~(\ref{Wet}), it has been discretized, using
the Ito prescription, in a one-dimensional lattice, using $1$ and
$0.001$ as discretization space and time meshes, respectively.
Systems sizes run from $L=50$ to $L=2000$, reaching up to $2 \times 10^8$
iteration time steps in the longest runs. Averages are performed over
$100$ to $10000$ runs (up to $2 \times 10^6$ in some cases).  Depending on
the values explored, different universality classes have been found:

i) {\bf MN:} Setting $b=c=D=\sigma^2=1$, we observe a
transition around $a_{\mathrm{mn}} \approx -0.01$ in the MN class
\cite{MN}.  We have also verified that the asymptotic properties do
not depend on parameter values as long as $b>0$.

ii) {\bf First order transition:} Changing the sign of $b$ (setting it
to $-3$), while keeping the rest of the parameters as in the previous
case, we find a discontinuous transition at $a_{\mathrm{ft}} =
1.85(1)$.  In this regime we recover all the phenomenology reported in
Ref.~\cite{Lisboa-wet}: an apparent first order phase transition, a
broad region of phase coexistence, triangular patterns in the
interface representation, etc.  However, by enlarging the value of $c$
the attractive well becomes less deep (see Fig.~\ref{Fig1}), the
discontinuity of the transition is reduced, and it becomes more and
more difficult to clearly establish the order of the transition.
Indeed, following a claim by Hinrichsen \cite{1?}  it could well be
the case that this discontinuous transition is only a transient
effect, asymptotically crossing-over to DP for very large system sizes.

\begin{figure}
\centerline{\psfig{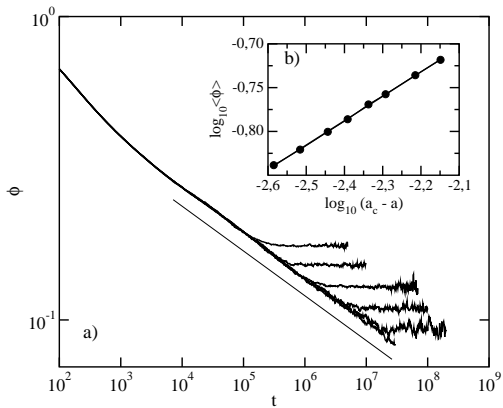}}
\caption{a) Log-log plot of the order parameter time evolution (averaged over
  surviving, pinned, runs) at the critical point, for system sizes
  (top to bottom) $L=25, 50, 100, 200$, $400$, and $1000$.  Both the
  main slope, $\theta=0.159(5)$, and the scaling of saturation values,
  $\beta/\nu = 0.245(15)$ reveal DP-scaling behavior. b) Log-log plot of
  the average order parameter as a function of $a_c -a$. The full line
  is a fit to the form 
$\langle \phi \rangle \sim (a_c - a)^\beta$, with an exponent
  $\beta=0.28(1)$, also compatible with DP.}
\label{Fig2}
\end{figure}
\begin{figure}
\centerline{\psfig{file=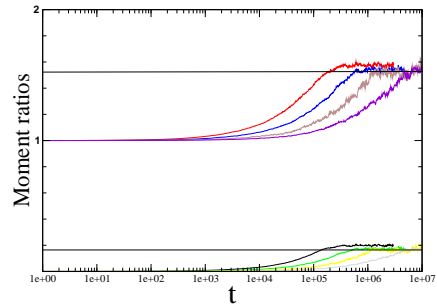,width=6cm,angle=-90}}
\caption{Time evolution of the moment ratios reported in the text
  for $L=25,50,100$ and $200$ at the critical point.  Note the
  convergence for large sizes to the expected DP values (straight
  lines).}
\label{Fig3}
\end{figure}


iii) {\bf DP:} In order to weaken the mechanism leading to a first
order transition---the mechanism responsible for destabilizing
depinned patches \cite{haye}---we consider a shallow well, $c=4$,
reduce the diffusion constant to $D=0.1$, and keep $b=-3$ and $\sigma=1$.
In Fig~\ref{Fig2}a) we show how the order parameter decays at
criticality ($a_c=0.1865(3)$) with a DP exponent $\theta \approx 0.16$ for
$L=1000$, and also how it saturates for smaller system sizes.  From
the scaling of the saturation values at criticality, we estimate
$\beta/\nu = 0.245(15)$, in good agreement with the DP value
$\beta/\nu=0.252...$ \cite{HGM}. As an independent check of universality,
we have also studied the dependence of $\langle \phi \rangle$ on $a$ at a fixed
system size $L=1000$. A fit to the form $\langle \phi \rangle \sim (a_c - a)^\beta$
provides the independent estimates $a_c=0.1871(5)$ and $\beta=0.28(1)$,
again compatible with DP, see Fig.~\ref{Fig2}b).  To further confirm
DP scaling, we have computed different moment and cumulant ratios at
criticality.  It has been recently shown that moment and cumulant
ratios are universal at the DP fixed point, and they are not strongly
affected by finite size effects. In particular, we have measured
$r_1=(m_2-m_1^2)/m_1^2$ (analogous to Binder's cumulant for these
transitions) and $r_2 = m_4/m_2^2$, where $m_i$ is the $i$-th moment.
In Fig.~\ref{Fig3} we show their time evolution at criticality for
different system sizes.  Observe the convergence to the extrapolated
DP values for infinite lattices $r_1\approx 0.173$ and $r_2 \approx 1.554$
\cite{moments}. Slight deviations from the values reported in
Ref.~\cite{moments} are due to lack of precision in the determination
of the critical point.

We have also measured a different order parameter: the density of
lattice sites with $h$ smaller than a certain reference
threshold--- {\it i. e.} the number of sites pinned at the attractive wall.  We
have checked that it scales as $\langle \phi \rangle$ (DP) at criticality,
independently of the arbitrary threshold.  Also, as a complementary
test to verify the generation of DP scaling for $b<0$, we have studied
the nature of fluctuations. In particular, we determine the
distribution of the output order parameter, $P(\langle \phi(t+1)\rangle)$, after
one iteration as a function of the input order parameter value $\langle
\phi(t) \rangle$.  It is not difficult to verify that its variance is
proportional to $\sqrt{\phi(t)}$, as expected for the DP class
\cite{HGM}.  In other words, the presence of an attractive wall
generates a DP-like noise $\sqrt{\phi(x,t)} \eta$, to be included in
Eq.~(\ref{MN}).  While in the absence of such a wall the leading
fluctuations are those intrinsic to the interface (linear in $\phi$) the
attractive wall generates a new noise term, related to the fluctuation
of the number of sites pinned by the potential well. This noise
dominates the asymptotic behavior, driving it from MN-type to DP.  Let us
remark that even though the transition is DP-like in this regime, it
has some first-order flavor. Indeed, as stressed before, there is a
finite phase coexistence interval $[a^*,a_c]$, within which pinned and
depinned interfaces coexist with a phenomenology very similar to
spatio-temporal-intermittency \cite{Lisboa-wet}.  For $a<a^*$
($a>a_c$) only the pinned (depinned) phase is
stable. A qualitatively similar ST have been found for CMLs 
and the term {\it fuzzy transition} has been coined to describe 
it \cite{firenze1}.

 From the theoretical side, we have performed a renormalization group 
perturbative calculation of Eq.(\ref{Wet}) along the same lines followed 
to study Eq.(\ref{MN}) \cite{MN}.
We find a weak coupling trivial fixed point (above $d=2$), 
and runaway trajectories above some separatrix analogously 
to what happens for Eq.(\ref{MN}) \cite{MN} (in particular, in $d=1$ these
are the only  solutions). The possible
existence of a DP strong-coupling fixed point cannot therefore be settled 
within this perturbative approach. 
This calls for new theoretical studies (see \cite{Ginelli}).

To sum up, all the observed universal features of synchronizing CML
are well reproduced by a simple Langevin equation, Eq.(\ref{Wet}),
including a potential with up to  quartic terms,
diffusion, and multiplicative noise, that is equivalent to a KPZ
equation with a bounding upper wall.  A MN transition is found for
$b>0$ (non-attractive wall), while for $b<0$ (attractive wall) first
order or DP transitions are observed depending upon parameter values.
In the MN case the scaling is controlled by interface fluctuations,
while in the second case it is dominated by fluctuations of the number
of sites pinned by the attracting wall (generating DP-type of noise).
This second mechanism has been shown to exhibit DP-scaling (probably, 
DP with many absorbing states \cite{Droz,HGM}) or,
alternatively, to induce (apparently) discontinuous transitions; 
in both cases, the transition is ``fuzzy'': there is a finite 
coexistence region between
synchronized and unsynchronized phases. 
In this way we have constructed a general framework to study ST in
spatially-extended systems and, in particular, in coupled map
lattices, successfully reproducing all their universal critical
features.  It would be very interesting to derive analytically the
Reggeon field theory Langevin equation describing DP \cite{HGM} from
Eq.~(\ref{MN}) in the adequate parameter range, and to develop
clean-cut criteria establishing the nature of the transition as a
function of bare parameter values (i. e. the full phase diagram).
These issues will be addressed in a future publication.

\begin{acknowledgments}
  We acknowledge very useful comments and discussions with R.
Livi, and also with F. de los Santos, A. Torcini,
  A. Politi, P. Grassberger, R. Dickman, and D. Mukamel, as well as
  financial support from the Spanish MCyT (FEDER) under project BFM2001-2841.
  R.P.-S. acknowledges also financial support from MCyT.
\end{acknowledgments}


\begin{thebibliography}{29}
\bibitem{ir} To be precise we should replace $b$ by its renormalized value.
\end{thebibliography}
\end{document}